\documentclass[aps,pra,twocolumn,amsmath,amssymb,nofootinbib,superscriptaddress]{revtex4}

\newcommand{\ket}[1]{|#1\rangle}

\usepackage[dvips]{graphicx}
\usepackage{mathrsfs}

\begin{document}

\bibliographystyle{apsrev}

\title{Noise thresholds for entanglement purification}

\author{Peter P. Rohde}
\email[]{rohde@physics.uq.edu.au}
\homepage{http://www.physics.uq.edu.au/people/rohde/}
\affiliation{Centre for Quantum Computer Technology, Department of Physics\\ University of Queensland, Brisbane, QLD 4072, Australia}

\date{\today}

\frenchspacing

\begin{abstract}
We consider the effects of gate noise on the operation of an entanglement purification protocol. We characterize the performance of the protocol by two measures, the minimum purifiable input state fidelity, and the maximum output state fidelity. Both these measures are a function of gate error rate. For sufficiently large gate error rate these two measures converge, defining a threshold on gate error rates. Numerically, we estimate this threshold to be $9\times 10^{-2}$, which is achievable with many present day experimental architectures.
\end{abstract}

\maketitle

\begin{figure}[!htb]
\includegraphics[width=0.6\columnwidth]{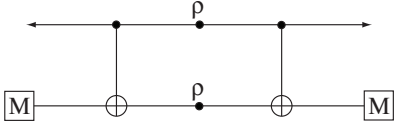}
\caption{Entanglement purification using two copies of a low fidelity state, $\hat\rho$, and two {\sc CNOT} gates (we have left out some local operations). Time flows from the center outwards. When the two measurement results are equal, we keep the resulting state, which will be of higher fidelity. If they disagree the state is discarded.} \label{fig:purification}
\end{figure}

Entanglement purification allows us to prepare an entangled state of higher fidelity from multiple copies of a lower fidelity entangled state. This has applications in quantum cryptography, quantum teleportation, quantum communication and quantum error correction. In Fig. \ref{fig:purification} we show the archetypal entanglement purification protocol by Bennett et al. \cite{bib:Bennett96}. In this paper we will focus on this protocol for its simplicity\footnote{Note that there is a vast array of other entanglement purification protocols, including ones for stabilizer states \cite{bib:Glancy06}, graph states \cite{bib:Goyal06, bib:Kay06, bib:Kruszynska06}, and continuous variable systems \cite{bib:Duan00}. Additionally, previous authors have considered the performance of such protocols under various noise models \cite{bib:Kay06b, bib:RohdeRalphMunro06}.}. Here two low fidelity copies of the $\ket\Psi = (\ket{01}+\ket{10})/\sqrt{2}$ Bell state are purified into a single higher fidelity one.

In this paper we consider the effects of noisy {\sc CNOT} gates on the operation of this protocol. See Refs. \cite{bib:BrielgelDur98,bib:DurBriegel99} for comparable analytic derivations\footnote{Also see Ref. \cite{bib:Aschauer02} for related work on secure communication in the presence of noisy gates.}. We model noisy {\sc CNOT}'s by following them with a generalized depolarizing channel that, with probability $p_\mathrm{gate}$, depolarizes both qubits acted upon by the gate,
\begin{equation}
\mathcal{E}(\hat\rho) = (1-p)\hat\rho + p\,\hat{I}\otimes\hat{I}.
\end{equation}

We simulate the system numerically by taking two copies of an initial state $\hat\rho$, and propagating it through the circuit from Fig. \ref{fig:purification}. We then make two identical copies of the output state and repeat the procedure. At each step we calculate the fidelity between the output state and the ideal Bell state $\ket\Psi$.

In Fig. \ref{fig:trajectories} we plot the fidelity of randomly chosen input states\footnote{We choose states by following ideal Bell pairs with depolarizing channels of random strength.} against the number of iterations of the entanglement purification protocol.
\begin{figure}[!htb]
\includegraphics[width=\columnwidth]{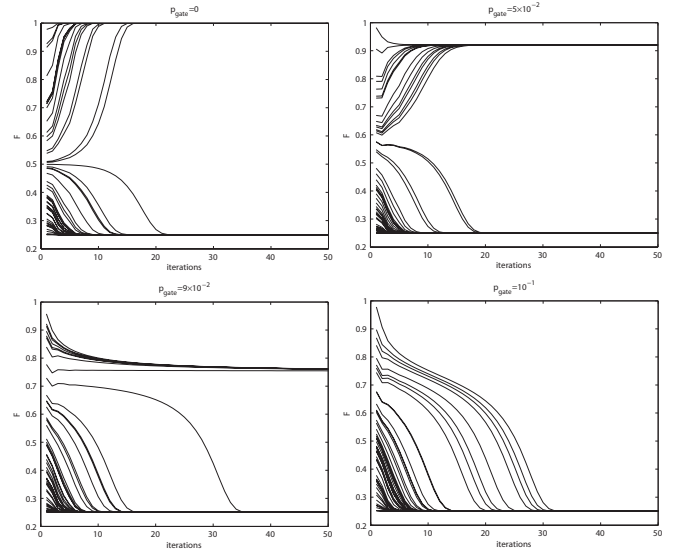}
\caption{Trajectories for random initial states under iterative application of the entanglement purification protocol.} \label{fig:trajectories}
\end{figure}

First consider the $p_\mathrm{gate}=0$ plot. Notice the divergence of the trajectories around $F=0.5$. For input states with $F>0.5$ the fidelity asymptotically approaches unity with sufficient rounds of the protocol. On the other hand, input states with $F<0.5$ are degraded by the protocol. This is consistent with the known fidelity bound for this protocol \cite{bib:Bennett96}.

Next consider the plot with some gate noise, $p_\mathrm{gate} = 5\times 10^{-2}$. The divergence point has shifted up from $F=0.5$ to $F\approx 0.6$. Thus, the minimum fidelity requirement for input states to be purifiable is higher. We label this value $F_\mathrm{min}$. Additionally, the asymptotic performance of the protocol, which we label $F_\infty$, has dropped to $F_\infty\approx0.92$.

In the next inset, $p_\mathrm{gate}=9\times 10^{-2}$, $F_\mathrm{min}$ and $F_\infty$ have converged. This represents a threshold on gate noise rate for entanglement purification to be possible. We label this $p_\mathrm{th}$. Beyond this threshold, as is shown in the fourth inset, the protocol is unable to purify any input states.

$F_\mathrm{min}$ has a direct physical interpretation as the minimum fidelity an input state must have for it to be purifiable. Similarly, $F_\infty$ has the interpretation as the maximum achievable fidelity of the output state.

In Fig. \ref{fig:trajectories} we plot $F_\mathrm{min}$ and $F_\infty$ against gate error rate. Notice the two lines converge at $p_\mathrm{gate}\approx 9\times 10^{-2}$, which we previously observed corresponds to the point at which the protocol fails to purify any input states. Thus, $p_\mathrm{th}\approx 9\times 10^{-2}$ (cf. Ref. \cite{bib:Kay06b} where upper bounds on noise rates in entanglement purification were derived analytically). For $p_\mathrm{gate}>p_\mathrm{th}$ these two line cross over and the protocol degrades the fidelity of input states.
\begin{figure}
\includegraphics[width=0.6\columnwidth]{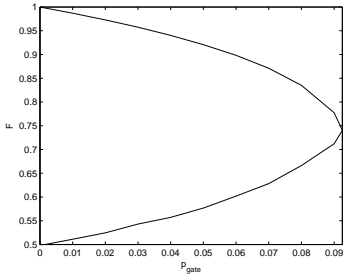}
\caption{(Upper line) Asymptotic fidelity, $F_\infty$. (Lower line) Minimum input state  fidelity, $F_\mathrm{min}$.} \label{fig:trajectories}
\end{figure}

In summary, we have considered the effects of gate noise on the operation of a well known entanglement purification protocol. Our results indicate a theoretical noise threshold of $p_\mathrm{th}=9\times 10^{-2}$. Experimental implementations will need to exhibit noise rates well below this threshold for meaningful entanglement purification to be possible. Furthermore, even below this threshold, the maximum achievable fidelity of purified states will be upper bounded as a function of gate noise rates. Thankfully, this threshold is relatively high and achievable with many present day technologies. As we have considered just a single protocol, it remains an open question as to what noise thresholds apply to other purification schemes.

\begin{acknowledgments}
We thank Henry Haselgrove, Alastair Kay, Andreas Winter and Timothy Ralph for helpful discussions. Simulations were carried out with the \emph{Quack!} package \cite{bib:RohdeHomepage}. This work was supported by the Australian Research Council and Queensland State Government. We acknowledge partial support by the DTO-funded U.S. Army Research Office Contract No. W911NF-05-0397.

\end{acknowledgments}

\bibliography{paper}

\end{document}